# Structural Controllability of Multi-Agent Networks: Robustness against Simultaneous Failures

Mohammad Amin Rahimian, Amir G. Aghdam

*Department of Electrical and Computer Engineering, Concordia University, 1455 de Maisonneuve Blvd. West, Montréal, Québec, H3G 1M8, Canada*

**Abstract**

In this paper, structural controllability of a leader-follower multi-agent system with multiple leaders is studied from a graph-theoretic point of view. The problem of preservation of structural controllability under simultaneous failures in both the communication links and the agents is investigated. The effects of the loss of agents and communication links on the controllability of an information flow graph are previously studied. In this work, the corresponding results are exploited to introduce some useful indices and importance measures that help characterize and quantify the role of individual links and agents in the controllability of the overall network. Existing results are then extended by considering the effects of losses in both links and agents at the same time. To this end, the concepts of joint $(r, s)-$controllability and joint $t-$controllability are introduced as quantitative measures of reliability for a multi-agent system, and their important properties are investigated. Lastly, the class of jointly critical digraphs are introduced and it is stated that if a digraph is jointly critical, then joint $t-$controllability is a necessary and sufficient condition for remaining controllable following the failure of any set of links and agents, with cardinality less than $t$. Various examples are exploited throughout the paper to elaborate on the analytical findings.

## 1 Introduction

The past decade has seen a growing interest in the control of multi-agent networks [1,2]. This type of system consists of a group of dynamic agents which interact according to a given information flow topology [1,2]. Distributed and cooperative control of these networked dynamic systems has found applications in emerging areas such as formation control of satellite clusters and motion coordination of robots [3,4]. An important class of multi-agent systems is the one with leader-follower architecture [5]. Various problems related to the control of leader-follower multi-agent systems include connectivity, containment, consensus, and flocking [6,7].

The problem of controllability in a Laplacian-based leader-follower multi-agent system with consensus-like interaction rules is first formulated by Tanner [8], where necessary and sufficient conditions for controllability are presented in terms of eigenvectors and eigenvalues of a sub-matrix of the graph Laplacian corresponding to the follower nodes. The importance of a graph theoretic characterization of controllability is also pointed out in [8]. The authors in [9] propose a sufficient condition for controllability of the network that is based on the null-spaces of the leader and follower incidence matrices. The condition is then restated in terms of the first homology of the network graph and its quotient over follower nodes. In a different attempt, Rahmani and Mesbahi [10] use Tanner's results to establish a relation between the notion of graph symmetry and the system theoretic concept of controllability by stating that symmetry with respect to follower nodes results in the uncontrollability of the network. The work [11] applies the concept of controllability to a network represented by a weighted directed graph and provides an interpretation for the controllability matrix in terms of the gains of fixed-length paths originated from the input node. Further results by Ji and Egerstedt [12] show that the existence of a common eigenvalue between the Laplacians of the original graph and the graph corresponding to the follower nodes is a necessary and sufficient condition for uncontrollability. This result is subsequently used to develop a sufficient condition based on the graph-theoretic concept of equitable partitions [13], which is then refined into necessary and sufficient conditions using relaxed equitable partitions [14]. Moreover, while the results relating graph symmetry to uncontrollability in [13] are shown to be explicable using equitable partitions, the relaxed equitable partitions in [14] provide a graph-theoretic characterization of the controllable subspace in the network. Another recent result shows that a multi-agent system is controllable if and only if no eigenvector of the graph Laplacian takes a zero value on the elements corresponding to the leaders [15]. Other challenging scenarios involving dynamic topologies and time-delay have also been investigated in the literature. For example, it is shown in [16] that switching between fixed uncontrollable topologies does not necessarily lead to an uncontrollable network.

⋆ This work has been supported by the Natural Sciences and Engineering Research Council of Canada (NSERC) under grant STPGP-364892-08.

The papers [17–19] approach the problem of link and agent failures by deriving graphical conditions for the preservation of structural controllability in the face of such failures. While existing results on the controllability of multi-agent systems provide an important measure of reliability of network to faults, they cannot handle the important problem of simultaneous failures of communication links and agents. The chief aim of this paper is to expand on the results of [17–20] by considering the case when communication links and agents in the network are both prone to failure. Moreover, the concepts of link and agent controllability degrees introduced in [17] and [19] are exploited to provide quantitative measures for the importance of individual links and agents in preserving the controllability of the overall digraph. In order to quantify the resilience of a network against multiple simultaneous failures, the notions of joint $(r, s)-$controllability and joint $t-$controllability are proposed and the latter is shown to be computable in polynomial-time. Next, a class of digraphs are investigated, for which joint $t-$controllability is a necessary and sufficient condition for remaining controllable after the failure of any set of links and agents with size less than $t$.

Identification and characterization of key link-points are very important for the reliable control of multi-agent networks. Furthermore, the comparative study of the importance of individual agents in the network is key to the design of reliable fault-tolerant multi-agent systems, as it provides guidelines on where to focus the recovery operations and which agents to prioritize in the case of a network-wide failure. The results are therefore of both theoretical and practical interest. On the other hand, the study of simultaneous failures is important in light of the fact that in real-world multi-agent systems, some faults can affect part of the network, containing a number of links and agents. This type of failure in multi-agent systems, where terrain properties or hardware faults disable a number of agents and limit the ability of others to communicate, motivates the study of controllability under simultaneous failure of links and agents in this work.

The remainder of this paper is organized as follows. Section 2 gives some preliminaries on sets and graph theory, and also reviews some results from [17] and [19]. The tools and concepts introduced in this section are then used in Section 3 to characterize and quantify the importance of every link and agent in the controllability of the overall digraph. In Section 4, first the notion of joint controllability is offered for the characterization of simultaneous failures in the network, and then the class of jointly critical digraphs are introduced and their useful properties are pointed out. The analytical results of Sections 3 and 4 are illustrated and discussed using some examples throughout the text, and concluding remarks are provided in Section 5.

## 2 Preliminaries and Notation

Throughout the paper, $\mathbb{N}$ denotes the set of all natural numbers, and $\mathbb{N}_k$ is the set of integers $\{1, 2, \ldots, k\}$. Furthermore, $\mathbb{R}$ denotes the set of all real numbers, $\mathbb{W} = \mathbb{N} \cup \{0\}$, and any other set is represented by a curved capital letter. The cardinality of a set $\mathscr{X}$ is denoted by $|\mathscr{X}|$ and in the strict mathematical sense, for two sets $\mathscr{X}$ and $\mathscr{Y}$ the inclusion symbols $\subset$ and $\subseteq$ are used interchangeably, while the latter emphasizes the possibility of the trivial inclusion $\mathscr{X} = \mathscr{Y}$ for some special configurations or problem scenarios. The difference of two sets $\mathscr{X}$ and $\mathscr{Y}$ is defined as $\mathscr{X} \backslash \mathscr{Y} = \{x | x \in \mathscr{X} \wedge x \notin \mathscr{Y}\}$. Moreover, $\mathscr{X}$ and $\mathscr{Y}$ are called disjoint if $\mathscr{X} \cap \mathscr{Y} = \varnothing$.

### 2.1 *Directed Information Flow Graph of a Multi-Agent System and its Controllability*

A directed graph or digraph is defined as an ordered pair of sets $(\mathscr{V}, \mathscr{E})$, where $\mathscr{V} = \{\nu_1, \ldots, \nu_n\}$ is the set of vertices and $\mathscr{E} \subseteq \mathscr{V} \times \mathscr{V}$ is the set of directed edges. In the graphical representation, each edge $\epsilon := (\tau, \nu) \in \mathscr{E}$ is denoted by a directed arc from the vertex $\tau \in \mathscr{V}$ to vertex $\nu \in \mathscr{V}$. Vertices $\nu$ and $\tau$ are referred to as the head and tail of the edge $\epsilon$, respectively. Notice that the definition of $\mathscr{E}$ does not allow for the existence of parallel arcs in the graphical representation of digraph $\mathscr{G} = (\mathscr{V}, \mathscr{E})$. Therefore, two edges that share the same pair of head and tail are identical. Given a set of vertices $\mathscr{X} \subset \mathscr{V}$, the set of all edges for which the tails belong to $\mathscr{X}$ but the heads do not, is termed the out-cut of $\mathscr{X}$, and is denoted by $\partial^+_{\mathscr{G}} \mathscr{X} \subset \mathscr{E}$. The cardinality of $\partial^+_{\mathscr{G}} \mathscr{X}$ is called the out-degree of $\mathscr{X}$, and is characterized as $d^+_{\mathscr{G}} \mathscr{X} = |\partial^+_{\mathscr{G}} \mathscr{X}|$. Similarly, the set of all edges for which the heads belong to $\mathscr{X}$ but the tails do not, is termed the in-cut of $\mathscr{X}$, and is denoted by $\partial^-_{\mathscr{G}} \mathscr{X} \subset \mathscr{E}$. Given an integer $k \in \mathbb{N}_{n-2}$, a set $\{\alpha_1, \alpha_2, \ldots, \alpha_k\} = \mathbb{N}_k$ and two vertices $\tau, \nu \in \mathscr{V}$, a sequence of distinct edges of the form $(\tau, \nu_{\alpha_1}), (\nu_{\alpha_1}, \nu_{\alpha_2}), \ldots, (\nu_{\alpha_{k-1}}, \nu_{\alpha_k})(\nu_{\alpha_k}, \nu)$ is called a $\tau\nu$ *path* if for any two edges $(\bar{\tau}, \bar{\nu}), (\hat{\tau}, \hat{\nu})$ of this sequence, $\bar{\nu} \neq \hat{\nu} \longleftrightarrow \bar{\tau} \neq \hat{\tau}$. For any $\mathscr{R} \subset \mathscr{V}$, a $\tau\nu$ path is called $\mathscr{R}-$rooted if $\tau \in \mathscr{R}$. The set $\mathscr{R}$ associated with an $\mathscr{R}-$rooted $\tau\nu$ path is referred to as the root-set, and a vertex $\nu \in \mathscr{V} \backslash \mathscr{R}$ is called reachable from the root-set $\mathscr{R}$ if there exists an $\mathscr{R}-$rooted $\tau\nu$ path, for some $\tau \in \mathscr{R}$. Two distinct $\tau\nu$ paths are called edge-disjoint if they do not share any edges. Two edge-disjoint $\tau\nu$ paths are called disjoint if $\tau$ and $\nu$ are the only vertices that are common to both of them.

Consider a team of $n$ single integrator agents given by:

$$\dot{x}_i(t) = u_i(t), \; i \in \mathbb{N}_n, \tag{1}$$

where the first $n-m$ agents are followers, and the last $m$ agents are leaders, with the following control inputs:

$$u_i(t) = \begin{cases} u_{ext}^i(t), & i \in \mathbb{N}_n \backslash \mathbb{N}_{n-m} \quad \text{(2a)} \\ \sum_{j \in \mathbb{N}_n} \alpha_{ij} x_j(t), & i \in \mathbb{N}_{n-m} \quad \text{(2b)} \end{cases}$$

where $\alpha_{ij} \in \mathbb{R}$ and $\alpha_{ii} \neq 0$ in (2b). Note that the leaders are influenced by external control inputs, whereas the followers are governed by a control law which is the linear combination of the states of neighboring agents as given by (2b). The interaction structure between the agents in (1) can be described by a directed information flow graph $\mathscr{G} = (\mathscr{V}, \mathscr{E})$, where each vertex represents an agent, and a directed edge from vertex $\nu_j$ to vertex $\nu_i$ indicates that $x_j(t)$ is transmitted to agent $i$ and $\alpha_{ij} \neq 0$ in (2b). Moreover, the condition $\alpha_{ii} \neq 0$ in (2b) implies the existence of a self-loop on each follower vertex of $\mathscr{G}$; however, the self-loops are omitted to simplify the graphical representations. In a digraph corresponding to a leader-follower multi-agent system, the root-set $\mathscr{R}$ consists of all leaders, and by assumption $|\mathscr{R}| = m$. The state of each agent $x_i(t)$ is its absolute position w.r.t. an inertial reference frame, and the agent dynamics is assumed to be decoupled along each axis of the frame.

**Remark 1** *Consider a leader-follower multi-agent system represented by the information flow digraph $\mathscr{G} = (\mathscr{V}, \mathscr{E})$ with the root-set $\mathscr{R}$. The control laws in* (2) *imply that no edges enter the root-set, i.e. $\partial_{\mathscr{G}}^{-}\mathscr{R} = \varnothing$.*

**Definition 1** *The information flow digraph $\mathscr{G}$ corresponding to the leader-follower multi-agent system* (1) *is called controllable if one can choose the non-zero coefficients $\alpha_{ij}$ in* (2b) *such that by properly moving the leaders, the followers would assume any desired configuration in an arbitrary time $T > 0$.*

The above definition of controllability, where the choices of non-zero parameters are arbitrary, is closely related to the study of controllability for linear structured systems [21,22]. The following theorem, borrowed from [19], provides a necessary and sufficient condition for the controllability of an information flow digraph as defined above, and is fundamental to all controllability results that follow.

**Theorem 1** *The information flow digraph $\mathscr{G} = (\mathscr{V}, \mathscr{E})$ with the root-set $\mathscr{R} \subset \mathscr{V}$ is controllable if and only if every vertex $\nu \in \mathscr{V} \backslash \mathscr{R}$ is reachable from the root-set $\mathscr{R}$.*

The next subsection summarizes the main results of [17] and [19], upon which Sections 3 and 4 expand.

### *2.2 Link and Agent Controllability Degrees*

Link and agent controllability degrees, defined below, provide quantitative insight into the reliability of a leader-follower multi-agent system in the face of agent and link failures and are known to be computable in polynomial-time, as investigated in [17] and [19] for a single leader and multiple leaders, respectively. A conceptually related issue is the fault tolerance of networks and the connectivity of their interconnection digraphs, as discussed in Section 1.5 of [23].

**Definition 2** *An information flow digraph $\mathscr{G} = (\mathscr{V}, \mathscr{E})$ with the root-set $\mathscr{R} \subset \mathscr{V}$ is said to be $p-$*link controllable *if $p$ is the largest number such that the controllability of the digraph is preserved after removing any group of at most $p-1$ edges. Moreover, a minimal set of $p$ edges whose removal makes $\mathscr{G}$ uncontrollable is referred to as a* critical link-set *and is denoted by $\mathscr{C}_p \subset \mathscr{E}$. A link is said to be* critical *if it belongs to a critical link-set and* uncritical *otherwise. The number $p$ is referred to as the* link controllability degree *of the digraph $\mathscr{G}$ w.r.t. the root-set $\mathscr{R}$, and is denoted by $lc(\mathscr{G};\mathscr{R})$.*

**Definition 3** *An information flow digraph $\mathscr{G} = (\mathscr{V}, \mathscr{E})$ with the root-set $\mathscr{R} \subset \mathscr{V}$ is said to be $q-$*agent controllable *if $q$ is the largest number such that the controllability of the digraph is preserved after removing any group of at most $q-1$ non-root vertices. Moreover, a minimal set of $q$ non-root vertices whose removal makes $\mathscr{G}$ uncontrollable is referred to as a* critical agent-set*, and is denoted by $\mathscr{C}_q \subseteq \mathscr{V} \backslash \mathscr{R}$. An agent is said to be* critical *if it belongs to a critical agent-set, and* uncritical *otherwise. The number $q$ is referred to as the* agent controllability degree *of the digraph $\mathscr{G}$ w.r.t. the root-set $\mathscr{R}$ and is denoted by $ac(\mathscr{G};\mathscr{R})$. Furthermore, for any $\nu \in \mathscr{V} \backslash \mathscr{R}$, the minimum number of vertices of $\mathscr{G}$ whose removal makes the vertex $\nu$ unreachable from the set $\mathscr{R}$ is denoted by $ac(\mathscr{G}, \nu; \mathscr{R})$.*

**Remark 2** *An information flow digraph $\mathscr{G} = (\mathscr{V}, \mathscr{E})$ with the root-set $\mathscr{R} \subset \mathscr{V}$ is not controllable if and only if $ac(\mathscr{G};\mathscr{R}) = lc(\mathscr{G};\mathscr{R}) = 0$. Moreover, for such a digraph all links and agents are assumed to be critical.*

**Remark 3** *Let $\mathscr{G}$, $\mathscr{V}$ and $\mathscr{R}$ be given as in Definition 3. For all $\nu \in \mathscr{V} \backslash \mathscr{R}$, if $\partial_{\mathscr{G}}^{+}\mathscr{R} \cap \partial_{\mathscr{G}}^{-}\{\nu\} \neq \varnothing$, then the follower agent $\nu$ remains reachable from the root-set $\mathscr{R}$ after the removal of any set of follower agents that does not include $\nu$. For such an agent $\nu$, the relation $ac(\mathscr{G}, \nu; \mathscr{R}) = |\mathscr{V}| - |\mathscr{R}|$ holds true.*

In the next section, the results of [24] and [25] are summarized, which provide quantitative measures for the importance of individual links and agents to the overall controllability of the network. A relevant discussion on the robustness of control law against link failure and the subsequent classification of links can be found in [26].

## 3 Importance of Individual Links and Agents to Network Controllability

According to Definition 2, the links in a multi-agent network can be categorized as critical and uncritical ones, based on their importance. Now to compare the role of two uncritical links, the designer can consider the resultant increase in the number of critical links in the network that is due to the removal of each uncritical link. This so-called *link controllability index* provides a means to determine which uncritical links take precedence in terms of their importance in the network. The next definition uses the notion of agent controllability degree (Definition 3) to introduce a second measure of importance, namely the *agent controllability index*, that would apply to both *critical* and *uncritical* links. Similarly to the link controllability index, the higher the agent controllability index of a link, the more important its role in the preservation of controllability throughout the network.

**Definition 4** *Consider an information flow digraph* $\mathscr{G} = (\mathscr{V}, \mathscr{E})$ *with the root-set* $\mathscr{R}$*. Let* $\epsilon \in \mathscr{E}$ *be an arbitrary edge in* $\mathscr{G}$*, and* $\mathscr{G}_1 = (\mathscr{V}, \mathscr{E}\backslash\{\epsilon\})$*. The* agent controllability index *of the edge* $\epsilon$ *is defined as* $\rho(\mathscr{G}, \epsilon; \mathscr{R}) := ac(\mathscr{G}; \mathscr{R}) - ac(\mathscr{G}_1; \mathscr{R})$*.*

**Lemma 1** *Consider an information flow digraph* $\mathscr{G} = (\mathscr{V}, \mathscr{E})$ *with the root-set* $\mathscr{R}$*. Let* $\epsilon \in \mathscr{E}\backslash\partial^+_{\mathscr{G}}\mathscr{R}$ *be an arbitrary edge in* $\mathscr{G}$ *whose head is the vertex* $\nu \in \mathscr{V}\backslash\mathscr{R}$*. If* $\mathscr{G}_1 = (\mathscr{V}, \mathscr{E}\backslash\{\epsilon\})$*, then* $\rho(\mathscr{G}, \epsilon; \mathscr{R}) = ac(\mathscr{G}, \nu; \mathscr{R}) - ac(\mathscr{G}_1, \nu; \mathscr{R})$*.*

**Proof.** The proof follows from the fact that the removal of the edge $\epsilon \in \mathscr{E}\backslash\partial^+_{\mathscr{G}}\mathscr{R}$ can affect the agent controllability degree of digraph $\mathscr{G}$ only by altering the available paths that connect the root-set $\mathscr{R}$ to the head vertex $\nu$, through other agents in the network. On the other hand, if $\epsilon \in \partial^+_{\mathscr{G}}\mathscr{R}$, then $\epsilon$ provides a direct path from $\mathscr{R}$ to $\nu$ which does not include any other agent in the network, and according to Remark 3, $ac(\mathscr{G}, \nu; \mathscr{R}) = |\mathscr{V}| - |\mathscr{R}|$, regardless of the available paths. ■

**Theorem 2** *Consider an information flow digraph* $\mathscr{G} = (\mathscr{V}, \mathscr{E})$ *with the root-set* $\mathscr{R}$*. If there exist a vertex* $\nu \in \mathscr{V}$ *and an edge* $\epsilon$ *such that* $\{\epsilon\} \subseteq \partial^+_{\mathscr{G}}\mathscr{R} \cap \partial^-_{\mathscr{G}}\{\nu\} \neq \varnothing$*, then* $\forall \hat{\epsilon} \in \partial^-_{\mathscr{G}}\{\nu\}\backslash\{\epsilon\}$*,* $\rho(\mathscr{G}, \hat{\epsilon}; \mathscr{R}) = 0$*.*

**Proof.** Let $\hat{\mathscr{G}} = (\mathscr{V}, \mathscr{E}\backslash\{\hat{\epsilon}\})$. For the case where $\hat{\epsilon} \notin \partial^+_{\mathscr{G}}\mathscr{R} \cap \partial^-_{\mathscr{G}}\{\nu\}$, the proof follows from Remark 3 and Lemma 1 upon noting that since the edge $\epsilon$ connects the vertex $\nu$ directly to the root-set, $ac(\mathscr{G}, \nu; \mathscr{R}) = ac(\hat{\mathscr{G}}, \nu; \mathscr{R}) = |\mathscr{V}| - |\mathscr{R}|$, and hence $ac(\mathscr{G}, \nu; \mathscr{R}) - ac(\hat{\mathscr{G}}, \nu; \mathscr{R}) = 0$. On the other hand, if $\hat{\epsilon} \in \partial^+_{\mathscr{G}}\mathscr{R} \cap \partial^-_{\mathscr{G}}\{\nu\}$, then the proof follows from the fact that both $\epsilon$ and $\hat{\epsilon}$ are providing a direct connection from the root-set $\mathscr{R}$ to their common head vertex $\nu$ that does not rely on any other follower agent in the network. Therefore, as long as this direct connection from the root-set $\mathscr{R}$ to the head vertex $\nu$ exists, removing either one of the edges $\epsilon$ or $\hat{\epsilon}$ would not impact the agent controllability degree of digraph $\mathscr{G}$. Hence, if $\{\epsilon, \hat{\epsilon}\} \subseteq \partial^+_{\mathscr{G}}\mathscr{R} \cap \partial^-_{\mathscr{G}}\{\nu\}$, then $\rho(\mathscr{G}, \epsilon; \mathscr{R}) = \rho(\mathscr{G}, \hat{\epsilon}; \mathscr{R}) = 0$. ■

Theorem 2 facilitates the characterization of the agent controllability index for those edges whose heads are directly connected to the root-set. In the special case that there exist multiple edges connecting the root-set to a vertex, Theorem 2 reduces to the following corollary.

**Corollary 1** *Given an information flow digraph* $\mathscr{G} = (\mathscr{V}, \mathscr{E})$ *with the root-set* $\mathscr{R}$ *and a vertex* $\nu \in \mathscr{V}\backslash\mathscr{R}$*, if* $|\partial^+_{\mathscr{G}}\mathscr{R} \cap \partial^-_{\mathscr{G}}\{\nu\}| > 1$*, then* $\forall \epsilon \in \partial^+_{\mathscr{G}}\mathscr{R} \cap \partial^-_{\mathscr{G}}\{\nu\}$*,* $\rho(\mathscr{G}, \epsilon; \mathscr{R}) = 0$*.*

The next theorem provides a full characterization of the agent controllability index for those edges whose heads are not directly connected to the root-set.

**Theorem 3** *Consider an information flow digraph* $\mathscr{G} = (\mathscr{V}, \mathscr{E})$ *with the root-set* $\mathscr{R}$*, and let* $\tau, \nu \in \mathscr{V}\backslash\mathscr{R}$ *be two vertices in* $\mathscr{G}$ *such that* $\epsilon := (\tau, \nu) \in \mathscr{E}\backslash\partial^+_{\mathscr{G}}\mathscr{R}$ *and* $\partial^-_{\mathscr{G}}\{\nu\} \cap \partial^+_{\mathscr{G}}\mathscr{R} = \varnothing$*. If* $\rho(\mathscr{G}, \epsilon; \mathscr{R}) \neq 0$*, then* $\rho(\mathscr{G}, \epsilon; \mathscr{R}) = 1$ *and there exists a critical agent-set* $\mathscr{C}_q \subset \mathscr{V}\backslash\mathscr{R}$ *of* $\mathscr{G}$ *such that* $\tau \in \mathscr{C}_q$*.*

**Proof.** Let $\mathscr{G}_1 = (\mathscr{V}, \mathscr{E}\backslash\{\epsilon\})$ and consider a critical agent-set $\mathscr{C}^1_q \subset \mathscr{V}\backslash\mathscr{R}$ of $\mathscr{G}_1$. If $\tau \in \mathscr{C}^1_q$ or $\nu \in \mathscr{C}^1_q$, then $\rho(\mathscr{G}, \epsilon; \mathscr{R}) = 0$, because $\mathscr{C}^1_q$ is then a critical agent-set of $\mathscr{G}$ as well. Since $\rho(\mathscr{G}, \epsilon; \mathscr{R}) \neq 0$, it is true that $\tau \notin \mathscr{C}^1_q$. The proof now follows upon noting that $\mathscr{C}_q = \mathscr{C}^1_q \cup \{\tau\}$ is a critical agent-set of $\mathscr{G}$ and $\tau \in \mathscr{C}_q$. ■

**Remark 4** *Theorems 2 and 3 address two mutually exclusive cases: the former applies to the incoming edges of the vertices that are directly connected to the root-set, while any other edge in the network is addressed by the latter.*

**Example 1** *Criticality and Agent Controllability Index.*

In all of the examples herein, nodes belonging to the root-set (leaders) are represented by dark vertices. The digraph $\mathscr{G}_1$ in Fig. 1(a) is 2−agent and 3−link controllable, and all of its links are critical. However, only for those links that belong to the out-cut of the root-set $\rho = 1$, and for the rest of the links $\rho = 0$.

Next, consider the digraph $\mathscr{G}_2$ of Fig. 1(b). Every dotted link in this digraph is critical (and vice versa) and it follows from Theorem 2 that those critical links which do not belong to the out-cut of the root-set have zero agent controllability index, while all other critical links have $\rho = 2$. On the other hand, the solid links are all uncritical with unity agent controllability index.

In what follows, the notions of agent and link controllability degrees are exploited to judge the importance of an agent as reflected through its outgoing links. The *agent* and *link criticality indices* are defined next, and the former is shown to distinguish between critical and uncritical agents based solely on their outgoing links.

**Definition 5** *Consider an information flow digraph* $\mathscr{G} = (\mathscr{V}, \mathscr{E})$ *with the root-set* $\mathscr{R}$*, and let* $\nu \in \mathscr{V}\backslash\mathscr{R}$

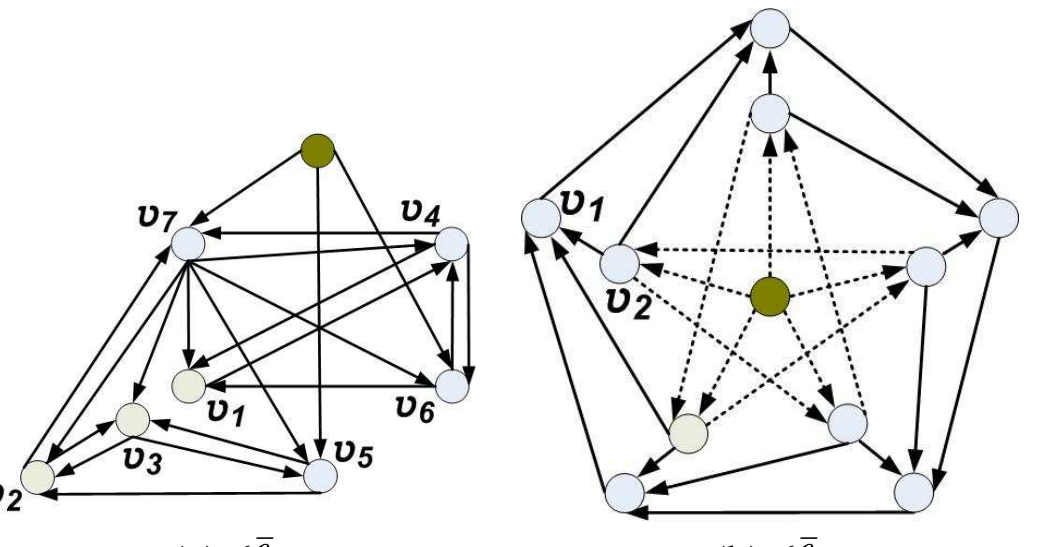

(a) $\bar{\mathscr{G}}_1$ (b) $\bar{\mathscr{G}}_2$

Fig. 1. (a) The digraph $\bar{\mathscr{G}}_1$, which is $2-$agent and $3-$link controllable. (b) The digraph $\bar{\mathscr{G}}_2$ of Example 1, for which critical and uncritical links are denoted by dotted and solid edges, respectively.

*be an arbitrary non-root vertex in $\mathscr{G}$. Let also $\mathscr{G}_1 = (\mathscr{V}, \mathscr{E}\backslash\partial^+_{\mathscr{G}}\{\nu\})$. The* agent criticality index *of vertex $\nu$ is denoted by $\delta(\mathscr{G}, \nu; \mathscr{R})$, and is given by $\delta(\mathscr{G}, \nu; \mathscr{R}) = ac(\mathscr{G}; \mathscr{R}) - ac(\mathscr{G}_1; \mathscr{R})$. In a similar manner, the* link criticality index *of vertex $\nu$ is characterized as $\theta(\mathscr{G}, \nu; \mathscr{R}) = lc(\mathscr{G}; \mathscr{R}) - lc(\mathscr{G}_1; \mathscr{R})$.*

**Lemma 2** *Given an information flow digraph $\mathscr{G} = (\mathscr{V}, \mathscr{E})$ with the root-set $\mathscr{R}$, if $ac(\mathscr{G}; \mathscr{R}) = |\mathscr{V}| - |\mathscr{R}|$, then $\forall \nu \in \mathscr{V}\backslash\mathscr{R}$, $\partial^-_{\mathscr{G}}\{\nu\} \cap \partial^-_{\mathscr{G}}\mathscr{R} \neq \varnothing$ and $\delta(\mathscr{G}, \nu; \mathscr{R}) = 0$. Conversely, if for all $\nu \in \mathscr{V}\backslash\mathscr{R}$, $\partial^-_{\mathscr{G}}\{\nu\} \cap \partial^-_{\mathscr{G}}\mathscr{R} \neq \varnothing$, then $ac(\mathscr{G}; \mathscr{R}) = |\mathscr{V}| - |\mathscr{R}|$ and $\forall \nu \in \mathscr{V}\backslash\mathscr{R}$, $\delta(\mathscr{G}, \nu; \mathscr{R}) = 0$.*

**Proof.** The proof for the first part follows by contradiction, since if $ac(\mathscr{G}; \mathscr{R}) = |\mathscr{V}| - |\mathscr{R}|$, then all follower agents are critical and $\mathscr{C}_q = \mathscr{V}\backslash\mathscr{R}$ is the only critical agent-set. Now, if there exists a vertex $\hat{\nu}$ that is not the head of a link belonging to the out-cut of the root-set $\mathscr{R}$, then the removal of the agent-set $\mathscr{C}_q\backslash\{\hat{\nu}\}$ will make $\hat{\nu}$ unreachable from the root-set $\mathscr{R}$, which is in contradiction with $\mathscr{C}_q$ being a critical agent-set. By the same token, to prove the converse suppose that $ac(\mathscr{G}; \mathscr{R}) < |\mathscr{V}| - |\mathscr{R}|$. Then there exist a critical agent-set $\mathscr{C}^1_q \subsetneqq \mathscr{V}\backslash\mathscr{R}$ and an agent $\tilde{\nu} \in \mathscr{V}\backslash\left(\mathscr{R} \cup \mathscr{C}^1_q\right)$ such that the removal of $\mathscr{C}^1_q$ will make $\tilde{\nu}$ unreachable from the root-set $\mathscr{R}$. This, however, is also a contradiction since $\tilde{\nu}$ is the head of a link belonging to the out-cut of the root-set $\mathscr{R}$. In both cases, the equality $\delta(\mathscr{G}, \nu; \mathscr{R}) = 0$ for all $\nu \in \mathscr{V}\backslash\mathscr{R}$ follows from the fact that $\forall \nu \in \mathscr{V}\backslash\mathscr{R}, ac(\mathscr{G}; \mathscr{R}) = ac(\mathscr{G}_1; \mathscr{R}) = |\mathscr{V}| - |\mathscr{R}|$, where $\mathscr{G}_1 = (\mathscr{V}, \mathscr{E}\backslash\partial^+_{\mathscr{G}}\{\nu\})$. ■

**Remark 5** *Digraph $\mathscr{G} = (\mathscr{V}, \mathscr{E})$ with root-set $\mathscr{R}$ for which $ac(\mathscr{G}; \mathscr{R}) = |\mathscr{V}| - |\mathscr{R}|$, corresponds to a pathological case where all agents are critical and they receive their "information" from the root-set "directly". In such a case, measures other than the agent criticality index are used to distinguish between the follower agents in terms of their significance in the network.*

**Theorem 4** *Given an information flow digraph $\mathscr{G} = (\mathscr{V}, \mathscr{E})$ with the root-set $\mathscr{R}$, suppose that $ac(\mathscr{G}; \mathscr{R}) < |\mathscr{V}| - |\mathscr{R}|$. For all $\nu \in \mathscr{V}\backslash\mathscr{R}$, $\delta(\mathscr{G}, \nu; \mathscr{R}) = 1$ if and only if $\nu$ is critical, and $\delta(\mathscr{G}, \nu; \mathscr{R}) = 0$ otherwise.*

**Proof.** Since $ac(\mathscr{G}; \mathscr{R}) < |\mathscr{V}| - |\mathscr{R}|$, the pathological case set forth in Lemma 2 does not apply. Let $\mathscr{G}_1 = (\mathscr{V}, \mathscr{E}\backslash\partial^+_{\mathscr{G}}\{\nu\})$, and suppose that $\mathscr{C}_q$ is an arbitrary critical agent-set of $\mathscr{G}$. The removal of $\mathscr{C}_q$ will make $\mathscr{G}_1$ uncontrollable, and if $\nu \in \mathscr{C}_q$, then $\mathscr{C}^1_q = \mathscr{C}_q\backslash\{\nu\}$ is a critical agent-set of $\mathscr{G}_1$. Hence, $\delta(\mathscr{G}, \nu; \mathscr{R}) = |\mathscr{C}_q| - |\mathscr{C}^1_q| = 1$. This proves that if $\nu$ is critical, then $\delta(\mathscr{G}, \nu; \mathscr{R}) = 1$. On the other hand, if $\nu$ is uncritical, then every critical agent-set of $\mathscr{G}$ is a critical agent-set of $\mathscr{G}_1$ and vice versa. Hence, $\delta(\mathscr{G}, \nu; \mathscr{R}) = ac(\mathscr{G}; \mathscr{R}) - ac(\mathscr{G}_1; \mathscr{R}) = 0$, which completes the proof. ■

**Remark 6** *Theorem 4 indicates that the criticality of an agent in any digraph (except for the pathological case described in Lemma 2 and Remark 5) is completely characterized by its outgoing links and through its agent criticality index given in Definition 5.*

The primary influence of an agent on the information flow structure is captured by the agent and link criticality indices $\delta$ and $\theta$, the latter being the more conclusive of the two. If two agents have the same criticality indices $\delta$ and $\theta$, then the one with a greater number of critical links amongst its out-going links, has a more prominent role, due to its effect in maintaining the critical links in the network. As another measure of importance, one can remove the uncritical links amongst the outgoing links of an agent, and consider the resultant increase in the number of critical links in the network. These last two effects are measured by the critical and uncritical link indices, respectively [24]. Together these indices offer a means for ordering and prioritizing agents in a network.

The following section and the subsequent definitions, lemmas, and theorems provide useful tools for investigating the impact of simultaneous link and agent failures on the controllability of an information flow digraph.

## 4 Joint Controllability

The concept of joint controllability degree parallels the notions of agent and link controllability degrees in Section 2, and facilitates the extension of the previous results to the cases where multiple simultaneous failures occur, affecting both links and agents in the network.

### 4.1 Joint Controllability Degree

**Definition 6** *An information flow digraph $\mathscr{G} = (\mathscr{V}, \mathscr{E})$ with the root-set $\mathscr{R} \subset \mathscr{V}$ is said to be* joint $(r, s)-$controllable *if it remains controllable in the case of simultaneous failure of any set of links of size $u \leqslant r$ and any set of non-root vertices of size $v \leqslant s$, where $u + v < r + s$ (note the strict inequality in the last expression).*

The next lemma follows immediately from Definitions 2, 3 and 6.

**Lemma 3** *The following statements hold:*

*a) If $\mathscr{G}$ is joint $(r,s)-$controllable, then for all $u \leqslant r$ and $v \leqslant s$, $\mathscr{G}$ is joint $(u,v)-$controllable.*
*b) If $\mathscr{G}$ is joint $(r,s)-$controllable, then $r \leqslant lc(\mathscr{G};\mathscr{R})$ and $s \leqslant ac(\mathscr{G};\mathscr{R})$.*
*c) If $\mathscr{G}$ is joint $(r,s)-$controllable and $lc(\mathscr{G};\mathscr{R}) = r$, then $s = 0$.*
*d) If $\mathscr{G}$ is joint $(r,s)-$controllable and $ac(\mathscr{G};\mathscr{R}) = s$, then $r = 0$.*

**Definition 7** *An information flow digraph $\mathscr{G} = (\mathscr{V},\mathscr{E})$ with the root-set $\mathscr{R} \subset \mathscr{V}$ is said to be* joint $t-$controllable *if $t$ is the largest number such that $\mathscr{G}$ is joint $(u,v)-$controllable for all $u+v \leqslant t$. Moreover, a minimal set of $r$ vertices and $s = t - r$ edges whose removal makes $\mathscr{G}$ uncontrollable is referred to as a* critical agent-link set*, and is denoted by $\mathscr{C}_{rs} \subset (\mathscr{V} \cup \mathscr{E})\backslash\mathscr{R}$. The number $t$ is called the* joint controllability degree *of the digraph $\mathscr{G}$ w.r.t. the root-set $\mathscr{R}$, and is denoted by $jc(\mathscr{G};\mathscr{R})$.*

From Definitions 6 and 7, it follows that a sufficient condition for the preservation of controllability in the face of simultaneous failures in links and agents is that the total number of failed links and agents is less than the joint controllability degree of the underlying information flow digraph.

**Theorem 5** *Given an information flow digraph $\mathscr{G} = (\mathscr{V},\mathscr{E})$ with the root-set $\mathscr{R}$, $jc(\mathscr{G};\mathscr{R}) = \min\{lc(\mathscr{G};\mathscr{R}), ac(\mathscr{G};\mathscr{R})\}$.*

**Proof.** The proof follows by contradiction. Let $lc(\mathscr{G};\mathscr{R}) = p$, $ac(\mathscr{G};\mathscr{R}) = q$ and $jc(\mathscr{G};\mathscr{R}) = t$, and suppose that $t < \min(p,q)$. From Definition 7, for some $\{r,s\} \subset \mathbb{N}_t$ and $r+s = t$, there exists a critical agent-link set $\mathscr{C}_{rs}$ which can be partitioned as $\mathscr{C}_{rs} = \mathscr{A} \cup \mathscr{L}$, where $\mathscr{A} \subset \mathscr{V}$, $\mathscr{L} \subset \mathscr{E}$, $|\mathscr{A}| = s$, and $|\mathscr{L}| = r$. Moreover, the removal of $\mathscr{C}_{rs}$ from $\mathscr{G}$ leads to an uncontrollable digraph $\mathscr{G}_1 = (\mathscr{V}_1,\mathscr{E}_1)$, where $\mathscr{V}_1 = \mathscr{V}\backslash\mathscr{A}$. Let $\mathscr{B}$ denote the set of agents corresponding to the heads of the links in $\mathscr{L}$. It follows that $|\mathscr{B}| \leqslant r$ and $\mathscr{B} \cap \mathscr{A} = \varnothing$, because otherwise the links whose heads belong to $\mathscr{B} \cap \mathscr{A}$ can be deleted from $\mathscr{C}_{rs}$ leading to a smaller agent-link set whose removal makes $\mathscr{G}$ uncontrollable, which contradicts with Definition 7 and $\mathscr{C}_{rs}$ being a critical agent-link set. Next, it follows from $\mathscr{B}\cap\mathscr{A} = \varnothing$ and $|\mathscr{B}| \leqslant r$ that $|\mathscr{A} \cup \mathscr{B}| \leqslant r+s = t < q$. Thus, the deletion of the agent-set $\mathscr{A} \cup \mathscr{B}$ from $\mathscr{G}$ leads to a controllable digraph $\mathscr{G}_2 = (\mathscr{V}_2,\mathscr{E}_2)$, where $\mathscr{V}_2 = \mathscr{V}_1\backslash\mathscr{B}$. This in turn implies that every agent belonging to $\mathscr{V}_1\backslash\mathscr{B}$ in the digraph $\mathscr{G}_1$ is reachable from $\mathscr{R}$ and vice versa. The latter converse statement follows from the fact that if there exists a vertex $\nu \in \mathscr{B}$ that is reachable from the root-set in $\mathscr{G}_1$, then the corresponding link in $\mathscr{L}$ whose head is $\nu$ can be deleted from $\mathscr{C}_{rs}$, leading to a contradictorily smaller critical agent-link set. Next note in the digraph $\mathscr{G}_1$ that since every vertex in $\mathscr{B}$ is unreachable from $\mathscr{R}$ while every vertex in $\mathscr{V}_1\backslash\mathscr{B}$ is reachable from $\mathscr{R}$, one should have $\partial^+_{\mathscr{G}_1}\mathscr{V}_1 \cap \partial^-_{\mathscr{G}_1}\mathscr{B} = \varnothing$ or equivalently $\partial^+_{\mathscr{G}}\mathscr{V}_1 \cap \partial^-_{\mathscr{G}}\mathscr{B} = \mathscr{L}$, because otherwise those vertices in $\mathscr{B}$ which are the heads of some links in $\partial^+_{\mathscr{G}_1}\mathscr{V}_1 \cap \partial^-_{\mathscr{G}_1}\mathscr{B}$ will be reachable from the root-set in $\mathscr{G}_1$. Now, consider an arbitrary vertex $\hat{\nu}$ belonging to the agent-set $\mathscr{B}$ in the digraph $\mathscr{G}$. Such a vertex is the head of some links in $\mathscr{L}$, say $l$ links, where $l \leqslant |\mathscr{L}|$. In addition to these $l$ links, $\hat{\nu}$ can only be the head of some links whose tails are either any of the $m \leqslant |\mathscr{L}| - l = r - l$ agents in $\mathscr{B}$ or any of the $s$ agents in $\mathscr{A}$. Hence, $|\partial^-_{\mathscr{G}}\{\hat{\nu}\}| \leqslant l + m + s \leqslant r + s$. On the other hand, the removal of $\partial^-_{\mathscr{G}}\{\hat{\nu}\}$ will make $\hat{\nu}$ unreachable from any vertex in $\mathscr{G}$, and therefore renders the digraph uncontrollable. This, however, is in contradiction with Definition 2 and the fact that $|\partial^-_{\mathscr{G}}\{\hat{\nu}\}| \leqslant r + s = t < p$. ■

The next definition and the theorem which follows, provide a mechanism to transform the problem of joint $t-$controllability of a given digraph into $q-$agent controllability of another digraph. This will, in turn, enable the multi-agent control system designer to take advantage of the polynomial-time algorithms developed in [17] and [19] for specifying the critical agent-link sets of a given digraph.

**Definition 8** *Given a digraph $\mathscr{G} = (\mathscr{V},\mathscr{E})$, replace every edge $\epsilon \in \mathscr{E}$ with two edges $\hat{\epsilon}_1$ and $\hat{\epsilon}_2$ in the same direction as $\epsilon$, and connect them through an intermediate vertex $\hat{\nu}_\epsilon$, termed a* black vertex. *The resulting digraph $\hat{\mathscr{G}} = (\hat{\mathscr{V}},\hat{\mathscr{E}})$ is called the* edge-duplicate *of $\mathscr{G}$. Every vertex of $\hat{\mathscr{G}}$ that is not a black vertex is referred to as a* white vertex.

**Remark 7** *Given a digraph $\mathscr{G} = (\mathscr{V},\mathscr{E})$ and its edge-duplicate $\hat{\mathscr{G}} = (\hat{\mathscr{V}},\hat{\mathscr{E}})$, the following equalities hold for the number of vertices and edges: $|\hat{\mathscr{V}}| = |\mathscr{V}| + |\mathscr{E}|$ and $|\hat{\mathscr{E}}| = 2|\mathscr{E}|$. Moreover, every white vertex $\hat{\nu}_\nu \in \hat{\mathscr{V}}$ corresponds to one vertex $\nu \in \mathscr{V}$ and every black vertex $\hat{\nu}_\epsilon \in \hat{\mathscr{V}}$ corresponds to one edge $\epsilon \in \mathscr{E}$. There exists a one-to-one correspondence between the sets $\mathscr{V} \cup \mathscr{E}$ and $\hat{\mathscr{V}}$.*

**Theorem 6** *Consider a digraph $\mathscr{G} = (\mathscr{V},\mathscr{E})$ with the root-set $\mathscr{R} \subset \mathscr{V}$ and its edge-duplicate $\hat{\mathscr{G}} = (\hat{\mathscr{V}},\hat{\mathscr{E}})$. The digraph $\mathscr{G}$ is joint $t-$controllable if and only if $\hat{\mathscr{G}}$ is $t-$agent controllable.*

**Proof.** The proof follows by construction, from the fact that Definition 8 specifies a bijection between the sets $\mathscr{V} \cup \mathscr{E}$ and $\hat{\mathscr{V}}$. Using this bijection, any critical agent-link set of $\mathscr{G}$ can be transformed into a critical agent-set of $\hat{\mathscr{G}}$ and vice versa. ■

In the next subsection, the important class of *jointly critical digraphs* is introduced and their role in characterizing robustness against simultaneous failures in both links and agents is highlighted.

### *4.2 Jointly Critical Digraphs*

The notion of agent controllability index from Definition 4 can be exploited to characterize and compare the relative susceptibility of digraphs with regard to agent or link failure.

Accordingly, in Lemmas 4 and 5, as well as Theorem 8 which follow, three classes of digraphs, termed as *agent-critical*, *link-critical* and *jointly critical*, are introduced and some of their important characteristics are pointed out.

**Lemma 4** *For an information flow digraph $\mathscr{G} = (\mathscr{V}, \mathscr{E})$ with root-set $\mathscr{R} \subset \mathscr{V}$, if $\forall \epsilon \in \partial^+_{\mathscr{G}}\mathscr{R}$, $\rho(\mathscr{G}, \epsilon; \mathscr{R}) = 1$, then $jc(\mathscr{G};\mathscr{R}) = ac(\mathscr{G};\mathscr{R})$. Such a digraph for which the aforementioned assumption holds will be referred to as* agent-critical.

**Proof.** Using Theorem 5, it suffices to introduce a set $\mathscr{A} \subset \mathscr{V}\backslash\mathscr{R}$ with the property $|\mathscr{A}| \leqslant lc(\mathscr{G};\mathscr{R})$, whose removal makes $\mathscr{G}$ uncontrollable. To this end, consider a solution $\mathscr{R} \subseteq \mathscr{X} \subset \mathscr{V}$ to the minimization problem $\min_{\mathscr{R}\subseteq\mathscr{X}\subset\mathscr{V}} d^+_{\mathscr{G}}\mathscr{X}$ in Theorem 3 of [19], which means that $|\partial^+_{\mathscr{G}}\mathscr{X}| = lc(\mathscr{G};\mathscr{R})$. Routine 1 utilizes $\partial^+_{\mathscr{G}}\mathscr{X}$ to generate one such set $\mathscr{A}$ with the desired characteristics. ■

**Routine 1**

1: $\mathscr{A} = \varnothing$
2: **for all** $(\tau, \nu) \in \partial^+_{\mathscr{G}}\mathscr{X}$ **do**
3:    **if** $\tau \notin \mathscr{R}$ **then**
4:       $\mathscr{A} = \mathscr{A} \cup \{\tau\}$
5:    **else**
6:       $\mathscr{A} = \mathscr{A} \cup \{\nu\}$
7:    **end if**
8: **end for**
9: **return** $\mathscr{A}$

**Remark 8** *When applying Routine 1 to an agent-critical digraph, it is notable that the assumption in Lemma 4 together with Corollary 1 ensures that step* 6 *will not be executed more than once for a given vertex $\nu$.*

Fig. 1(a) shows the case of an agent-critical digraph, for which the agent and link controllability degrees are given by $q = 2$ and $p = 3$, respectively, and they satisfy the relation $q \leqslant p$, as suggested by Lemma 4.

**Lemma 5** *Consider an information flow digraph $\mathscr{G} = (\mathscr{V}, \mathscr{E})$ with the root-set $\mathscr{R} \subset \mathscr{V}$. If there exists a critical agent-set $\mathscr{C}_q \subset \mathscr{V}\backslash\mathscr{R}$ of $\mathscr{G}$ such that $\forall \nu \in \mathscr{C}_q$, $\exists \epsilon \in \partial^+_{\mathscr{G}}\{\nu\}$ for which $\rho(\mathscr{G}, \epsilon; \mathscr{R}) \neq 0$, then $jc(\mathscr{G};\mathscr{R}) = lc(\mathscr{G};\mathscr{R})$. Such a digraph for which the aforementioned assumption holds will be referred to as* link-critical.

**Proof.** The proof follows by using Theorem 5 and introducing a set $\mathscr{B} \subset \mathscr{E}$ with the property $|\mathscr{B}| \leqslant ac(\mathscr{G};\mathscr{R})$, whose removal makes $\mathscr{G}$ uncontrollable. Let $\mathscr{C}_q$ be a critical agent-set satisfying the condition of Lemma 5. Routine 2 utilizes $\mathscr{C}_q$ to generate one such set $\mathscr{B}$ with the property that $\mathscr{B} \subset \mathscr{E}\backslash\partial^+_{\mathscr{G}}\mathscr{R}$ and $|\mathscr{B}| = |\mathscr{C}_q| = ac(\mathscr{G};\mathscr{R})$. ■

**Remark 9** *When applying Routine 2 to a link-critical digraph, it is notable that with $\mathscr{C}_q$ satisfying the conditions of Lemma 5, step* 5 *will be executed exactly once for every vertex $\nu \in \mathscr{C}_q$.*

**Routine 2**

1: $\mathscr{B} = \varnothing$
2: **for all** $\nu \in \mathscr{C}_q$ **do**
3:    **for all** $\epsilon \in \partial^+_{\mathscr{G}}\{\nu\}$ **do**
4:       **if** $\mathscr{B} \cap \partial^+_{\mathscr{G}}\{\nu\} = \varnothing$ **and** $\rho(\mathscr{G}, \epsilon; \mathscr{R}) = 1$ **then**
5:          $\mathscr{B} = \mathscr{B} \cup \{\epsilon\}$
6:       **end if**
7:    **end for**
8: **end for**
9: **return** $\mathscr{B}$

**Remark 10** *Using Theorem 3, it can be stated that digraph $\mathscr{G}$ in Lemma 5 is link-critical if there exists a critical agent-set $\mathscr{C}_q \subset \mathscr{V}\backslash\mathscr{R}$ of $\mathscr{G}$ such that $\forall \nu \in \mathscr{C}_q$, $\exists \epsilon \in \partial^+_{\mathscr{G}}\{\nu\}$ for which $\rho(\mathscr{G}, \epsilon; \mathscr{R}) = 1$.*

Digraph $\mathscr{G}_2$ in Fig. 1 is 3-agent and 2-link controllable. This digraph is link-critical, and satisfies the condition of Lemma 5.

**Theorem 7** *Consider a joint $(r, s)-$controllable information flow digraph $\mathscr{G} = (\mathscr{V}, \mathscr{E})$ with the root-set $\mathscr{R} \subset \mathscr{V}$. If $\mathscr{G}$ is agent-critical or link-critical, then $r + s \leqslant \max\{lc(\mathscr{G};\mathscr{R}), ac(\mathscr{G};\mathscr{R})\}$.*

**Proof.** According to the results of Lemmas 4 and 5, it suffices to prove that if $\mathscr{G}$ is agent-critical, then $r + s \leqslant lc(\mathscr{G};\mathscr{R})$, and if $\mathscr{G}$ is link-critical, then $r + s \leqslant ac(\mathscr{G};\mathscr{R})$. For an agent-critical digraph $\mathscr{G}$, consider a solution $\mathscr{R} \subseteq \mathscr{X} \subset \mathscr{V}$ to the minimization problem $\min_{\mathscr{R}\subseteq\mathscr{X}\subset\mathscr{V}} d^+_{\mathscr{G}}\mathscr{X}$. According to Theorem 3 of [19], the link controllability degree of $\mathscr{G}$ is equal to the out-degree of $\mathscr{X}$, i.e. $|\partial^+_{\mathscr{G}}\mathscr{X}| = lc(\mathscr{G};\mathscr{R})$, and it follows from part (b) of Lemma 3 that $r \leqslant lc(\mathscr{G};\mathscr{R})$. If $r = lc(\mathscr{G};\mathscr{R})$, then part (c) of Lemma 3 requires that $s = 0$, and hence the statement of the above theorem holds. If on the other hand $r < lc(\mathscr{G};\mathscr{R})$, then choose a set of edges $\mathscr{Z}_r \subset \mathscr{E}$ such that $\mathscr{Z}_r \subset \partial^+_{\mathscr{G}}\mathscr{X}$ and $|\mathscr{Z}_r| = r$. Use Routine 1 after replacing $\partial^+_{\mathscr{G}}\mathscr{X}$ with $\partial^+_{\mathscr{G}}\mathscr{X}\backslash\mathscr{Z}_r$ to generate a set $\mathscr{A} \subset \mathscr{V}$. Now, $\mathscr{A} \cup \mathscr{Z}_r$ is a set of $|\mathscr{A}|$ vertices and $|\mathscr{Z}_r| = r$ edges, for which $\mathscr{G}_{\mathscr{A},\mathscr{Z}_r} = (\mathscr{V}\backslash\mathscr{A}, \mathscr{E}\backslash\mathscr{Z}_r)$ is uncontrollable. However, $\mathscr{G}$ is joint $(r, s)-$controllable, which implies that $s \leqslant |\mathscr{A}|$. On the other hand, it follows from Routine 1 that $|\mathscr{A}| \leqslant |\partial^+_{\mathscr{G}}\mathscr{X}\backslash\mathscr{Z}_r|$. Hence $s \leqslant |\partial^+_{\mathscr{G}}\mathscr{X}\backslash\mathscr{Z}_r|$ or $s \leqslant lc(\mathscr{G};\mathscr{R}) - r$, which completes the proof for an agent-critical digraph. A similar argument can be utilized to prove the statement for the case where $\mathscr{G}$ is link-critical. From part (b) of Lemma 3, it is clear that $s \leqslant ac(\mathscr{G};\mathscr{R})$. Now, starting from a critical agent-set $\mathscr{C}_q$ that satisfies the condition of Lemma 5, select an arbitrary subset $\mathscr{Z}_s \subset \mathscr{C}_q$ of $s = |\mathscr{Z}_s|$ agents. If $s = ac(\mathscr{G};\mathscr{R})$, then part (d) of Lemma 3 requires that $r = 0$ and the statement for the link-critical case holds. If $s < ac(\mathscr{G};\mathscr{R})$, then applying Routine 2 to $\mathscr{C}_q\backslash\mathscr{Z}_s$ yields a set $\mathscr{B}$ of $|\mathscr{C}_q\backslash\mathscr{Z}_s| = ac(\mathscr{G};\mathscr{R}) - s = |\mathscr{B}|$ links, whose deletion together with deletion of the $s$ agents in $\mathscr{Z}_s$ will render $\mathscr{G}$ uncontrollable. The proof for a link-critical digraph $\mathscr{G}$ follows now upon the realization that since $\mathscr{G}$ is joint $(r, s)-$controllable, one should have $|\mathscr{B}| = ac(\mathscr{G};\mathscr{R}) - s \geqslant r$. ■

**Remark 11** *Using the pathological class of digraphs described in Lemma 2 and Remark 5, it is straightforward to construct a joint $(r,s)-$controllable digraph $\mathscr{G}$ with the root-set $\mathscr{R}$, such that $r+s > \max\{ac(\mathscr{G};\mathscr{R}), lc(\mathscr{G};\mathscr{R})\}$. Non-trivial counterexamples are also possible and important. The joint $(3,2)-$controllable digraph in Fig. 2 is $4-$link and $3-$agent controllable.*

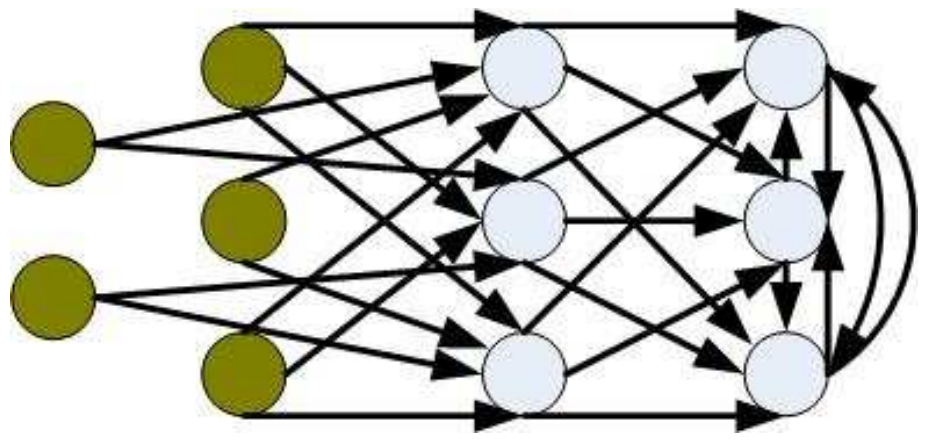

Fig. 2. An example of a joint $(r,s)-$controllable digraph $\mathscr{G}$ which does not satisfy the relation $r+s \leqslant \max\{ac(\mathscr{G};\mathscr{R}), lc(\mathscr{G};\mathscr{R})\}$.

Theorems 5 and 6 in [17] provide the following bounds on the number of edges $|\mathscr{E}|$ of a digraph $\mathscr{G} = (\mathscr{V},\mathscr{E})$: $|\mathscr{E}| \geqslant (|\mathscr{V}|-1)lc(\mathscr{G};\mathscr{R})$ and $|\mathscr{E}| \geqslant |\mathscr{V}| + ac(\mathscr{G};\mathscr{R}) - 2$. These two inequalities, together with Theorem 5 in the previous subsection, imply that $|\mathscr{E}| \geqslant \max\{(|\mathscr{V}|-1)\ jc(\mathscr{G};\mathscr{R}),\ |\mathscr{V}| + jc(\mathscr{G};\mathscr{R}) - 2\}$. In the same vein, Theorem 7 proved in this subsection, yields the following corollary, specifying some necessary conditions on the number of edges of an agent-critical or link-critical digraph, which is joint $(r,s)-$controllable. These results can be used in the design of reliable multi-agent control systems.

**Corollary 2** *Consider a joint $(r,s)-$controllable information flow digraph $\mathscr{G} = (\mathscr{V},\mathscr{E})$. The following statements hold: if $\mathscr{G}$ is agent-critical, then $|\mathscr{E}| \geqslant (|\mathscr{V}|-1)(r+s)$. Moreover, if $\mathscr{G}$ is link-critical, then $|\mathscr{E}| \geqslant |\mathscr{V}| + r + s - 2$.*

The next theorem and the remark that follows capture the significance of joint controllability degree for the so-called *jointly critical* digraphs.

**Theorem 8** *If an information flow digraph $\mathscr{G} = (\mathscr{V},\mathscr{E})$ with the root-set $\mathscr{R} \subset \mathscr{V}$ is both agent-critical and link-critical, then $jc(\mathscr{G};\mathscr{R}) = ac(\mathscr{G};\mathscr{R}) = lc(\mathscr{G};\mathscr{R})$. Moreover, for every $(r,s) \in \mathbb{W}\times\mathbb{W}$, the digraph $\mathscr{G}$ is joint $(r,s)-$controllable if and only if $r+s \leqslant jc(\mathscr{G};\mathscr{R})$. Such a digraph, which is both agent-critical and link-critical, will be referred to as* jointly critical.

**Proof.** From Lemmas 4 and 5, it is immediate that $jc(\mathscr{G};\mathscr{R}) = ac(\mathscr{G};\mathscr{R}) = lc(\mathscr{G};\mathscr{R})$. The rest of the proof, also, can be sketched as a combination of the proofs of Lemmas 4 and 5. Starting from a critical link-set $\mathscr{C}_p \subseteq \mathscr{E}$, one can use Routine 1 to transform any set of links $\mathscr{L} \subseteq \mathscr{C}_p$ into a set of agents $\mathscr{A}$, where $|\mathscr{A}| \leqslant |\mathscr{L}|$, such that the removal of $\mathscr{A}$ along with the links in $\mathscr{C}_p \backslash \mathscr{L}$ renders the digraph uncontrollable. Moreover, $|\mathscr{A}| = |\mathscr{L}|$ because the inequality $|\mathscr{A}| < |\mathscr{L}|$ contradicts the fact that $ac(\mathscr{G};\mathscr{R}) = lc(\mathscr{G};\mathscr{R}) = |\mathscr{C}_p|$. On the other hand, starting from a critical agent-set $\mathscr{C}_q \subseteq \mathscr{V}\backslash\mathscr{R}$ that satisfies the condition of Lemma 5, one can use Routine 2 to transform any set of agents $\mathscr{B} \subseteq \mathscr{C}_q$ into a set of $|\mathscr{B}|$ links, which if removed together with the agents in $\mathscr{C}_q \backslash \mathscr{A}$, then the digraph becomes uncontrollable. ■

**Remark 12** *If a digraph $\mathscr{G}$ with the joint controllability degree $t$ is jointly critical, then for all $(r,s) \in \mathbb{W}\times\mathbb{W}$ satisfying the inequality $r+s > t$, $\mathscr{G}$ is not joint $(r,s)-$controllable. Hence, the joint controllability degree alone completely characterizes the controllability preservation properties of the digraph $\mathscr{G}$. In other words, if the values of $(r,s) \in \mathbb{W}\times\mathbb{W}$ for which $\mathscr{G}$ is joint $(r,s)-$controllable are depicted as discrete points in the plane, then a pair of non-negative integers belongs to the jointly controllable set if and only if the corresponding point in the $(r,s)$-plane lies in the region $r+s \leqslant t$.*

Three special cases of interest are addressed in the sequel.

#### *4.2.1 Complete Digraphs*

As a special case, a digraph $\mathscr{G}_{cn} = (\mathscr{V}_{cn},\mathscr{E}_{cn})$ is called *complete* if $\mathscr{E}_{cn} = \mathscr{V}_{cn}\times\mathscr{V}_{cn}$. Select a vertex $r$ in a complete digraph as the root, and remove the $|\mathscr{V}_{cn}|-1$ edges headed by the vertex $r$. Then the resultant information flow digraph is $(|\mathscr{V}_{cn}|-1)-$link controllable [17]. This is the maximum value for the link controllability degree in an information flow digraph with $n = |\mathscr{V}_{cn}|$ vertices, because a complete digraph possesses the maximum possible number of edges per a given number of vertices. The following proposition suggests that the joint controllability degree of a complete information flow digraph $\mathscr{G}_{cn}$ is also $n-1$.

**Proposition 1** *Given a complete digraph $\mathscr{G}_{cn} = (\mathscr{V}_{cn},\mathscr{E}_{cn})$ with $|\mathscr{V}_{cn}| = n$, choose a vertex $r$ as the root and remove the $n-1$ edges which are headed by $r$. The resulting information flow digraph is jointly critical and joint $(n-1)-$controllable.*

**Proof.** The proof follows from the fact that $\mathscr{G}_{cn}$ has exactly $n-1$ disjoint $r\nu$ paths for every $\nu \in \mathscr{V}_{cn}\backslash\{r\}$. ■

**Remark 13** *It is to be noted that $jc(\mathscr{G}_{cn};\{r\}) = n-1$ is the highest attainable joint controllability degree for a digraph with $n$ vertices. This explains the desirable controllability preservation properties of the complete digraphs in the face of simultaneous link and agent failures.*

#### *4.2.2 Kautz Digraphs*

Kautz digraphs are introduced and discussed in Section 3.3 of [23]. Accordingly, a Kautz digraph $\mathscr{G}_k = (\mathscr{V}_k,\mathscr{E}_k)$ with $|\mathscr{V}_k| = n$ is given by:

$$\mathscr{V}_k = \{\nu_1,\ldots,\nu_n\}, \quad (3)$$
$$\mathscr{E}_k = \{(\nu_i,\nu_j) | i,j \in \mathbb{N}_n \wedge j \equiv (-id-\tau) \bmod n,\ \tau \in \mathbb{N}_d\},$$

for some $d \in \mathbb{N}\backslash\{1\}$ and $\kappa \in \mathbb{N}$, such that $d^{\kappa} + d^{\kappa-1} = n$. The following proposition gives the joint link controllability degree of an information flow digraph derived from a Kautz digraph.

**Proposition 2** *Consider a Kautz digraph $\mathscr{G}_k = (\mathscr{V}_k, \mathscr{E}_k)$ where $\mathscr{V}_k$ and $\mathscr{E}_k$ are given by* (3). *Choose a vertex $r$ as the root and remove all edges which are headed by $r$. The resulting information flow digraph is jointly critical and joint $d-$controllable.*

**Proof.** The proof follows upon noting that $\mathscr{G}_k$ has exactly $d$ disjoint $r\nu$ paths for every $\nu \in \mathscr{V}_k \backslash \{r\}$. ■

### 4.2.3 *Circulant Digraphs*

Circulant digraphs are introduced and discussed in Section 3.4.5 of [23]. Accordingly, a circulant digraph $\mathscr{G}_c = (\mathscr{V}_c, \mathscr{E}_c)$ with $|\mathscr{V}_c| = n$ is given by:

$$\begin{aligned} \mathscr{V}_c &= \{\nu_1, \ldots, \nu_n\}, \\ \mathscr{E}_c &= \{(\nu_i, \nu_j) | i, j \in \mathbb{N}_n \wedge j - i \equiv b \bmod n,\ b \in \mathscr{B}\}, \end{aligned} \quad (4)$$

for some $\mathscr{B} \subseteq \mathbb{N}_{n-1}$. Choose a vertex $r \in \mathscr{V}_c$ as the root, and remove every edge whose head is $r$. Then in the resulting information flow digraph $\mathscr{G}_c$, $lc(\mathscr{G}_c; \{r\}) = |\mathscr{B}|$. This is due to the fact that $\mathscr{G}_c$ has exactly $|\mathscr{B}|$ edge-disjoint $r\nu$ paths for every $\nu \in \mathscr{V}_c \backslash \{r\}$. For $|\mathscr{V}_i| = 5,\ i \in \mathbb{N}_3$, the choices of $\mathscr{B}_1 = \{1\}$, $\mathscr{B}_2 = \{1, n-1\}$, and $\mathscr{B}_3 = \{1, n-2\}$ correspond to a simple loop $\mathscr{G}_1 = (\mathscr{V}_1, \mathscr{E}_1)$, a distributed double-loop $\mathscr{G}_2 = (\mathscr{V}_2, \mathscr{E}_2)$, and a daisy chain loop $\mathscr{G}_3 = (\mathscr{V}_3, \mathscr{E}_3)$, respectively. These digraphs are introduced in Section 3.4.1 of [23], and they are depicted in Figs. 3(a)−(c). For a simple loop, the equality $lc(\mathscr{G}_1; \{r\}) = ac(\mathscr{G}_1; \{r\}) = jc(\mathscr{G}_1; \{r\}) = 1$ holds, while for the other two cases $lc(\mathscr{G}_i; \{r\}) = ac(\mathscr{G}_i; \{r\}) = jc(\mathscr{G}_i; \{r\}) = 2,\ i = 2, 3$. These three digraphs have the additional property that for any $r$, $s$ satisfying the inequality $r + s > jc(\mathscr{G}_i; \{r\}),\ i \in \mathbb{N}_3$, $\mathscr{G}_i$ is not joint $(r, s)-$controllable. Accordingly, the joint controllability degree alone completely characterizes the controllability preservation properties for $\mathscr{G}_i,\ i \in \mathbb{N}_3$. This is due to the fact that $\mathscr{G}_i,\ i \in \mathbb{N}_3$, are jointly critical.

On the other hand, for the circulant digraph $\mathscr{G}_4$ with $|\mathscr{V}_4| = 6$, $\mathscr{B}_4 = \{2, 3, 5\}$ and the uppermost vertex selected as the root $r$, the resulting information flow digraph, shown in Fig. 3(d), is $3-$link and $2-$agent controllable [17]. This digraph is neither agent-critical nor link-critical, and hence is not jointly critical. The joint controllability degree for $\mathscr{G}_4$ is 2, and unlike $\mathscr{G}_i,\ i \in \mathbb{N}_3$, $jc(\mathscr{G}_4; \{r\}) = 2$ does not proffer a full characterization of the controllability preservation properties for $\mathscr{G}_4$. Accordingly, $\mathscr{G}_4$ is joint $(r, s)-$controllable for $(r, s) \in \{(2, 1), (3, 0)\}$, although $r + s > jc(\mathscr{G}_4; \{r\})$. To clarify this point, let the values of $(r, s) \in \mathbb{W} \times \mathbb{W}$ for which $\mathscr{G}_i$ is joint $(r, s)-$controllable, $i \in \mathbb{N}_4$, be shown as discrete points in the plane. For $i \in \mathbb{N}_3$, the line $r + s = jc(\mathscr{G}_i; \{r\})$ divides the first quadrant of the $(r, s)$-plane into two regions, where for $r + s \leqslant jc(\mathscr{G}_i; \{r\})$, $\mathscr{G}_i$ is joint $(r, s)-$controllable and otherwise it is not. This is depicted in Fig. 4 for $\mathscr{G}_2$, $\mathscr{G}_3$ and $\mathscr{G}_4$. In the case of $\mathscr{G}_2$ and $\mathscr{G}_3$, the closed shaded region contains all pairs of integers belonging to the joint controllability set (the points associated with these pairs are shown by black circles). This property, however, does not hold for $\mathscr{G}_4$, where there exist two points above the line $r + s = jc(\mathscr{G}_4; \{r\})$ representing the pairs for which $\mathscr{G}_4$ is still joint $(r, s)-$controllable.

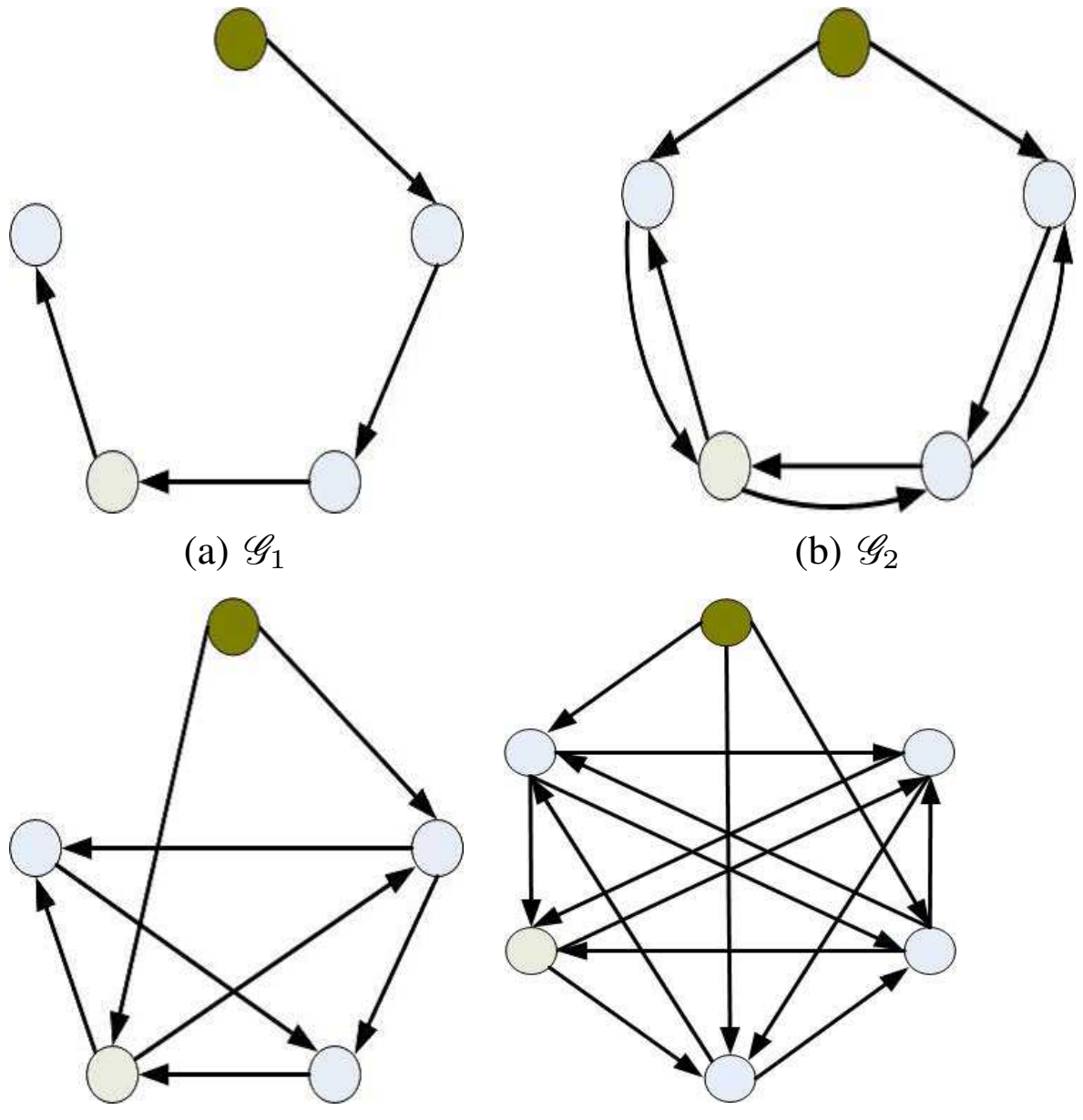

(a) $\mathscr{G}_1$ (b) $\mathscr{G}_2$ (c) $\mathscr{G}_3$ (d) $\mathscr{G}_4$

Fig. 3. Circulant digraphs

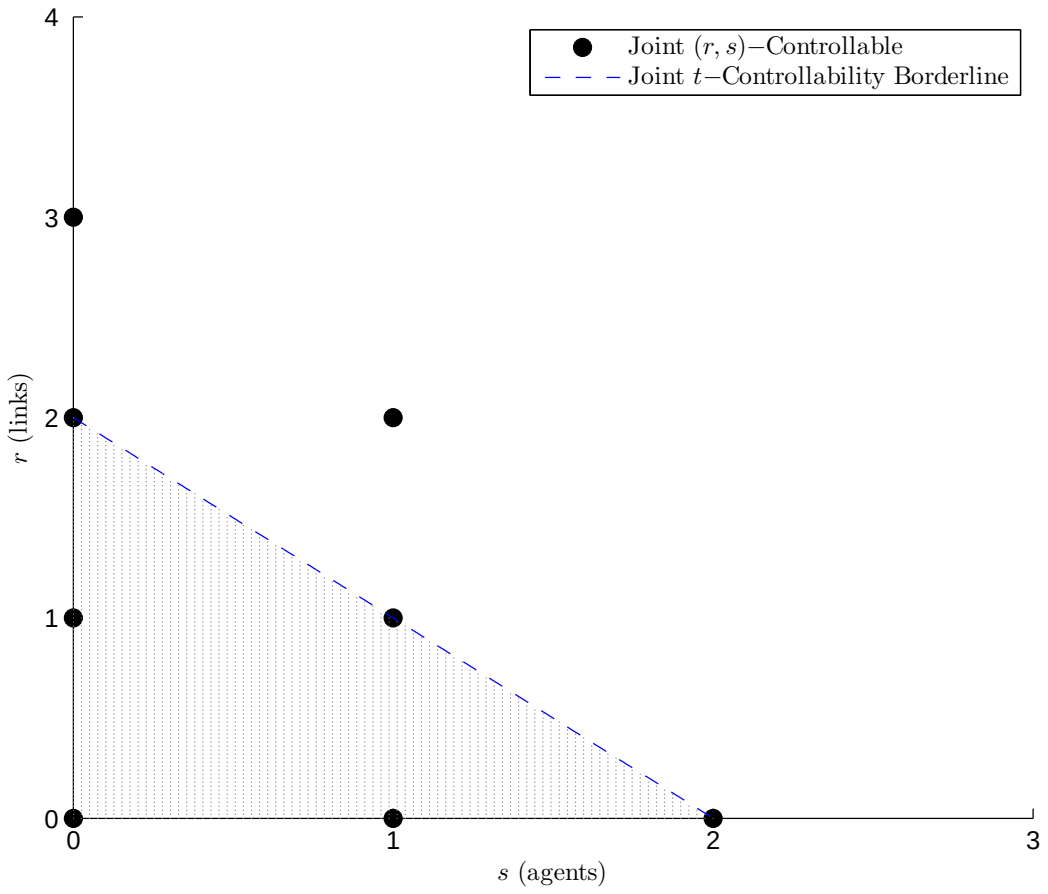

Fig. 4. The joint controllability of the circulant digraphs $\mathscr{G}_2$, $\mathscr{G}_3$ and $\mathscr{G}_4$ given in Figs. 3(b) to 3(d) is considered here. For $\mathscr{G}_2$ and $\mathscr{G}_3$, the shaded area contains all of the filled circles representing the pairs of integers that belong to their jointly controllable set; this property, however, does not hold true in the case of $\mathscr{G}_4$.

## 5 Conclusions

Structural controllability of a network of single-integrator agents with leader-follower architecture was investigated. The notions of agent controllability index, as well as agent and link criticality index were defined to characterize and quantify the importance of individual links and agents to the controllability of the overall network. The results provide the designer of a multi-agent system with useful means to evaluate (and enhance) the reliability of the network by deciding on which links and agents to prioritize for fault management and recovery operations.

In the next step, the concepts of joint $(r, s)-$controllability and joint $t-$controllability were proposed as quantitative measures of reliability in a multi-agent system subject to simultaneous failures of communication links and agents. It was noted that joint $t-$controllability is a *conservative* requirement which provides a *sufficient* condition for remaining controllable following the removal of any set of links and agents with size less than $t$. Nonetheless, for the important class of jointly critical digraphs, the joint controllability degree $t$ proffers a *necessary and sufficient* condition, which fully characterizes the controllability preservation properties of the digraph. By and large, a digraph remains controllable after the removal of any $u$ links and $v$ agents *if and only if* there exists a pair $(r, s) \in \mathbb{W} \times \mathbb{W}$ such that the digraph is joint $(r, s)-$controllable and $u \leqslant r \wedge v \leqslant s \wedge u + v < r + s$. However, the authors' ongoing research indicates that for some digraphs, which are neither agent nor link critical, determining *all* $(r, s)$ pairs for which the digraph is joint $(r, s)-$controllable may not be tractable in polynomial-time, and future research on this topic is of much interest. The presented results provide design guidelines for improving the network robustness against simultaneous failures of multiple links and agents. A number of examples were offered to elucidate the results.